# Decentralized Coordinated State Estimation in Integrated Transmission and Distribution Systems

Ying Zhang*, Yanbo Chen†, Jianhui Wang‡, Yue Meng*, and Tianqiao Zhao*

*Department of Interdisciplinary Science,
Brookhaven National Laboratory, Upton, NY 11973, USA
†Department of Electrical and Electronic Engineering,
North China Electric Power University, Beijing 102206, China
‡Department of Electrical and Computer Engineering,
Southern Methodist University, Dallas, TX 75205, USA

*Abstract*— Current transmission and distribution system states are mostly unobservable to each other, and state estimation is separately conducted in the two systems owing to the differences in network structures and analytical models. The large-scale integration of transmission and active distribution systems calls for an effective solution to global state estimation. Unlike existing independent state estimation methods on both levels of these systems, we propose a decentralized coordinated transmission and distribution system state estimation (C-TDSE) method. This method enables accurate monitoring of the integrated systems with a global reference in a decentralized manner and reconciles the mismatches of voltages and powers on boundaries of the systems. The comparative analysis on the integrated transmission and distribution systems points to improved estimation results relative to the independent state estimation methods.

## I. Introduction

State estimation as a critical tool converts redundant meter readings and other available information into an estimate of system states [1]. Historically, transmission system state estimation (TSSE) and distribution system state estimation (DSSE) have been studied separately as two subjects owing to significant differences in network structures and algorithmic procedures, and attract substantial works, reviewed as [2].

Distribution systems are physically coupled with transmission systems, forming integrated transmission and distribution (T&D) systems. These systems are separately managed by transmission system operators (TSOs) and distribution system operators (DSOs) [3]. The penetration of distributed generators (DGs) poses uncertainties and more stringent requirements for system operation [4]. Also, the uncertainty introduced to distribution systems may affect their top-level transmission grid. This indicates the limitation of this siloed system management mechanism, and instead, the T&D coordination brings the improvement of operational efficiency and economic benefits [6]. However, coordinated T&D state estimation for monitoring the integrated systems is difficult to run in a centralized manner, as TSOs and DSOs do not share the complete system models and measurement data due to privacy and storage burdens [1]. Besides, technical difficulties persist in centralized estimation methods for these T&D systems due to differences in system observability and topology structure [3]. Linear state estimators are developed in the transmission systems with the installation of phasor measurement units (PMUs), which provides global positioning system (GPS) synchronization [7], [8]. In comparison, limited meters in conventional supervisory control and data acquisition (SCADA) systems and pseudo-measurements with low accuracy are widely used in current DSSE methods [9].

Multi-area state estimation methods are proposed to monitor interconnected power systems in a distributed manner. These methods are applied to homogeneous systems, *e.g.*, transmission and transmission systems [7], [8], [10]–[13] or distribution and distribution systems [14], [15]. For instance, [12] introduces a multilevel and hierarchical state estimation paradigm, and three major levels in interconnected power systems, including local TSSE, multi-area TSSE, and regional multi-TSO state estimation are identified. This theoretical framework is further elaborated by considering various measurement configurations and specific requirements in these multi-area systems. At the distribution level, a limited number of meters are considered, resulting in poor system observability.

In interconnected T&D systems, heterogeneous decomposition is regarded as an efficient solution to coordinated T&D operation. It evolves a subject of active research since [3] develops a master-slave-splitting iterative method for solving the global power flow in integrated T&D systems. This paper proposes a decentralized coordinated transmission and distribution system state estimation (C-TDSE) algorithm in integrated T&D systems, considering the heterogeneous network structures and meter configurations in these systems. The proposed algorithm enables effective monitoring of the integrated T&D systems and a reduction of the mismatches of state variables at boundary nodes from multiple local estimators. The comparative analysis points to the improved estimation results relative to these independent state estimation methods. Also, such improvement is polished by limited data exchange and fast coordination on the boundaries of these T&D systems. The main contributions are summarized below.

- The decentralized C-TDSE method obtains the improved accuracy and low computational costs by limited information interchange in integrated T&D systems.
- The proposed method reduces the mismatches of powers on the boundaries of these systems and coordinates all the

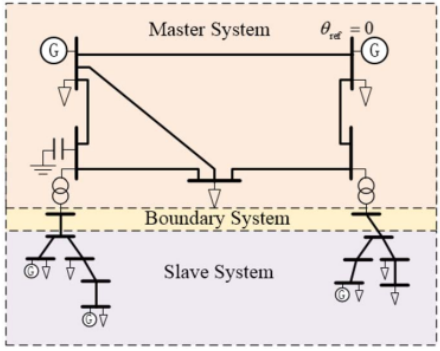

Fig.1. An integrated T&D system with a master-slave structure [6]

voltage phase angles with a global reference.

## II. STATE ESTIMATION AND INTEGRATED T&D SYSTEM

### A. State Estimation Theory

In standard state estimators [1], the relationship between measurements and state variables are depicted as:
$$z = h(x) + e \quad (1)$$
where $x \in \mathbb{R}^{n \times 1}$ denotes the state vector, and $z \in \mathbb{R}^{m \times 1}$ denotes the measurement vector; $h(x)$ is the measurement function vector about $x$; The measurement error vector $e$ follows Gaussian distributions as $e \sim N(0, R)$, where the measurement covariance matrix $R = diag[\sigma_1^2, \sigma_2^2, ..., \sigma_m^2]$.

The estimated state variables are obtained via a weighted least square (WLS) criterion that minimizes the sum of weighted measurement residuals $J$ as:
$$\hat{x} = \arg\min J = \arg\min r^T W r \quad (2)$$
where the weight matrix $W$ is chosen as the inverse of $R$, and $r = z - h(x)$ represents the measurement residual vector.

Let $\partial J/\partial x = 0$, and optimal estimated states are solved iteratively by the Gauss-Newton method. The estimation procedure terminates until each component of the vector $\Delta x$ at iteration $\tau$ is sufficiently small, i.e., $\Delta x < \varepsilon$:
$$H(x^{(\tau)})^T W H(x^{(\tau)}) \Delta x = H(x^{(\tau)})^T W [z - h(x^{(\tau)})] \quad (3)$$
$$x^{(\tau+1)} = x^{(\tau)} + \Delta x \quad (4)$$
where $H(x^{(\tau)})$ is the Jacobian matrix with respect to the states $x^{(\tau)}$, and $H(x^{(\tau)}) = \partial h(x^{(\tau)})/\partial x^{(\tau)}$.

The covariance matrix of the estimated states, $\mathrm{cov}(\hat{x})$, is used to quantify the estimation variances of these states and calculated by [12]
$$\mathrm{cov}(\hat{x}) = G^{-1} \quad (5)$$
where $\hat{x}$ is the final estimated state vector until the above iterative process terminates, and $G = H(\hat{x})^T W H(\hat{x})$ is the gain matrix of this estimator.

### B. Master-slave Structure of Integrated T&D Systems

Current TSOs and DSOs separately monitor and control the system operation via meters installed within their jurisdiction. The system states are unobservable to each other as TSSE and DSSE algorithms run separately. The master-slave structure for integrated T&D systems depicts the system characteristics efficiently and thus is widely used [6]. The integrated system is divided into a master system, boundary systems, and slave systems, shown as Fig. 1. The boundary system is composed of the substation between the T&D systems. A transmission system includes a master system and boundary systems that are connected to distribution systems, while a distribution system consists of a slave system and the boundary system connected with the transmission system. Three types of nodes are defined:
- Master Node: A node only belonging to the master system.
- Boundary Node: A substation node that is connected to transmission and distribution systems.
- Slave Node: A node only belonging to the slave system.

## III. PROPOSED ALGORITHM

Based on the master-slave structure, we present a C-TDSE algorithm for monitoring integrated T&D systems in a decentralized manner, including local state estimation phase, coordination phase, and update phase.

### A. Local TSSE

Transmission systems are assumed observable by only PMU measurements, resulting in a linear estimator owing to the increasing popularization of PMUs in transmission systems, e.g., [7], [8], and [12]. The real and imaginary parts of voltages at all nodes are chosen as state variables:
$$X_t = [V_{r1}, ..., V_{rN}, V_{x1}, ..., V_{xN}]^T \quad (6)$$
where $X_t \in \mathbb{R}^{2N \times 1}$ and $N$ is the number of all the nodes.

The measurements consist of the voltage and current phasors recorded by PMUs. Also, the local TSSE algorithm depicts the relationship between these state variables and measurements as
$$z_t = h_t(x^M, x_1^B, x_2^B, ..., x_f^B) + e_t \quad (7)$$
where $X_t$ is divided into two groups, $x^M$ and $x_i^B$ denote the states in the master system and the $i$th connected boundary system, and $i = 1, 2, ..., f$; $z_t$ denotes the PMU measurements, including the voltage and current phasors and $z_t = \begin{bmatrix} z_V \\ z_I \end{bmatrix}$, while $e_t$ denotes the measurement noises.

We express the measurement function as:
$$h_t(X_t) = \begin{bmatrix} I \\ Y \end{bmatrix} \cdot X_t = H_t X_t \quad (8)$$
where $I$ denotes the identity matrix relating the voltage phasors to the states, and $Y$ denotes the system admittance matrix.

The solution of this linear model is solved by
$$X_t = (H_t^T W_t H_t)^{-1} H_t^T W_t z_t \quad (9)$$
Based on (5), the covariance of these estimated variables is given by $\mathrm{cov}(\hat{X}_t) = G_t^{-1} = (H_t^T W_t H_t)^{-1}$.

### B. Local DSSE

SCADA systems are widely used to provide voltage magnitude and power measurements in current distribution systems. Also, a substation acts as a slack node, and the phase angle of this node is set as zero. Voltage magnitudes at all nodes and the voltage phase angles except that of the substation node are chosen as state variables [9]. Assume that the number of the distribution systems is $f$. The states $X_{d,i}$ in the $i$th distribution system are expressed as
$$X_{d,i} = [V_1, V_2, ..., V_{N_i}, \theta_2, \theta_3, ..., \theta_{N_i}]^T \quad (10)$$

TABLE I
JACOBIAN ELEMENTS IN THE COORDINATION PHASE

| Jacobian Elements | | $y_i = v_n$ | | $y_i = \alpha_n$ | |
|---|---|---|---|---|---|
| | | $n = k$ | $n \in \mathcal{N}(i)\backslash\{k\}$ | $n = k$ | $n \in \mathcal{N}(i)\backslash\{k\}$ |
| $\boldsymbol{H}^M$ | $\partial V_{rn}^M/\partial y_i$ | $cos\alpha_n$ | | $-\boldsymbol{v}_n sin\alpha_n$ | |
| | $\partial V_{xn}^M/\partial y_i$ | $sin\alpha_n$ | | $\boldsymbol{v}_n cos\alpha_n$ | |
| $\boldsymbol{H}^S$ | $\partial V_k^S/\partial y_i$ | 1 | 0 | 0 | 0 |
| $\boldsymbol{H}^B$ | $\partial f_{PMB}/\partial y_i$ | $f_{PMB}/v_k + g_{nk}v_k$ | $v_n(g_{nk}cos\alpha_{nk} + b_{nk}sin\alpha_{nk})$ | $f_{QMB} + b_{nk}\boldsymbol{v}_k^2$ | $v_k v_n(g_{nk}sin\alpha_{nk} - b_{nk}cos\alpha_{nk})$ |
| | $\partial f_{QMB}/\partial y_i$ | $f_{QMB}/v_k - b_{nk}v_k$ | $v_n(g_{nk}sin\alpha_{nk} - b_{nk}cos\alpha_{nk})$ | $-f_{PMB} + g_{nk}\boldsymbol{v}_k^2$ | $v_k v_n(-g_{nk}cos\alpha_{nk} - b_{nk}sin\alpha_{nk})$ |

where $\boldsymbol{X}_{d,i} \in \mathbb{R}^{(2N_i-1)\times 1}$, and $N_i$ is the number of nodes in the $i$th distribution system, $i = 1, 2, ..., f$.

The measurement functions are expressed as

$$\begin{cases} \boldsymbol{z}_{d,1} = \boldsymbol{h}_{d,1}\left(\boldsymbol{x}_1^{B'}, \boldsymbol{x}_1^S\right) + \boldsymbol{e}_{d,1} \\ \boldsymbol{z}_{d,2} = \boldsymbol{h}_{d,2}\left(\boldsymbol{x}_2^{B'}, \boldsymbol{x}_2^S\right) + \boldsymbol{e}_{d,2} \\ \vdots \\ \boldsymbol{z}_{d,f} = \boldsymbol{h}_{d,f}\left(\boldsymbol{x}_f^{B'}, \boldsymbol{x}_f^S\right) + \boldsymbol{e}_{d,f} \end{cases} \quad (11)$$

where $\boldsymbol{X}_{d,i}$ is divided into two sub-vectors, $\boldsymbol{x}_i^{B'}$ and $\boldsymbol{x}_i^S$, which denote the state vectors in the $i$th boundary system and its slave distribution system; $\boldsymbol{e}_{d,i}$ is the measurement noise vector, while $\boldsymbol{z}_{d,i}$ and $\boldsymbol{h}_{d,i}\left(\boldsymbol{x}_i^{B'}, \boldsymbol{x}_i^S\right)$ denote the corresponding measurements and measurement function.

The measurement vector includes the voltage magnitudes and power flows recorded by the SCADA systems and pseudo-measurements. The detailed formulation of $\boldsymbol{h}_{d,i}(\boldsymbol{X}_{d,i})$ in (11) and its Jacobian matrix can be found in [9]. The estimated states are obtained by the iterative procedure (3) and (4), where the initial state vector is $\boldsymbol{X}_{d,i}^{(0)} = [V_{s,i}, ..., V_{s,i}, 0, ..., 0]^T$, i.e., a flat start is used. $V_{s,i}$ denotes the voltage magnitude at the substation of boundary system $i$. The iterative tolerance is set as $\varepsilon_d = 10^{-4}$. The covariance matrix of this estimator is given by $\text{cov}\left(\widehat{\boldsymbol{X}}_{d,i}\right) = \boldsymbol{G}_{d,i}^{-1}$.

### C. Coordination Phase

A coordination phase in the integrated T&D systems is developed to 1) reconcile the mismatches of the estimated states at boundary nodes in the local phases, 2) estimate the phase angles of these slave systems in coordinated universal time (UTC), and 3) further refine the estimated states in the T&D systems. In this phase, we choose the state variables of boundary system $i$ as $\boldsymbol{y}_i = \{\boldsymbol{v}_n, \boldsymbol{\alpha}_n\}$, where $\boldsymbol{v}_n$ and $\boldsymbol{\alpha}_n$ denote the voltage magnitude and phase angle vectors at node $n$, and $n \in \mathcal{N}(i)$; $\mathcal{N}(i)$ denotes the set of boundary node $k$ in boundary system $i$ and the nodes connected to this node; $\boldsymbol{y}_i \in \mathbb{R}^{2M_i \times 1}$, and $M_i$ is the total number of the nodes in $\mathcal{N}(i)$.

The relationship between the state variables and these locally estimated states ideally holds in boundary system $i$:

$$\begin{cases} V_{rn}^M = \boldsymbol{v}_n \cos\boldsymbol{\alpha}_n & n \in \mathcal{N}(i) \\ V_{xn}^M = \boldsymbol{v}_n \sin\boldsymbol{\alpha}_n & n \in \mathcal{N}(i) \end{cases} \quad (12)$$

$$V_k^S = \boldsymbol{v}_n \quad n = k \quad (13)$$

where $V_k^S$ denotes the estimated voltage magnitude from $\boldsymbol{x}_i^B$ in the local DSSE phase, while $V_{rn}^M$ and $V_{xn}^M$ denote the real and imaginary parts of the estimated voltage at node $n$ from $\boldsymbol{x}_i^M$ and $\boldsymbol{x}_i^{B'}$ in the local TSSE phase. These estimates act as auxiliary measurements in the coordination phase.

The following power equations hold in boundary system $i$:

$$\begin{cases} P_i^S + P_i^B = f_{PMB}(\boldsymbol{y}_i) \\ Q_i^S + Q_i^B = f_{QMB}(\boldsymbol{y}_i) \end{cases} \quad (14)$$

where $P_i^S$ and $Q_i^S$ denote the active and reactive powers flowing in or out of the substation to the slave system; $P_i^B$ and $Q_i^B$ denote the sums of active and reactive power injection measurements at the boundary node.

The active and reactive power flows from the master system to the connected slave system via the boundary node $k$ are expressed by

$$f_{PMB}(\boldsymbol{y}_i) = \sum_{n \in \mathcal{N}(i)} v_n v_k (g_{nk}cos\alpha_{nk} + b_{nk}sin\alpha_{nk}) \quad (15)$$

$$f_{QMB}(\boldsymbol{y}_i) = \sum_{n \in \mathcal{N}(i)} v_n v_k (g_{nk}sin\alpha_{nk} - b_{nk}cos\alpha_{nk}) \quad (16)$$

where $\alpha_{nk}$ denotes the phase angle difference between boundary node $k$ and node $n$, and $\alpha_{nk} = \alpha_n - \alpha_k$; $g_{nk}$ and $b_{nk}$ represent the real and imaginary parts of the nodal admittance between nodes $n$ and $k$, $n \in \mathcal{N}(i)$.

Considering the estimation errors at the local stages and measurement noises, a nonlinear estimator is established in boundary system $i$:

$$\boldsymbol{z}_{y,i} = \boldsymbol{h}(\boldsymbol{y}_i) + \boldsymbol{e}_{y,i} \quad (17)$$

$$\partial J(\boldsymbol{y}_i)/\partial \boldsymbol{y}_i = \boldsymbol{H}(\boldsymbol{y}_i)^T \boldsymbol{W}_{y,i}[\boldsymbol{z}_{y,i} - \boldsymbol{h}(\boldsymbol{y}_i)] = \boldsymbol{0} \quad (18)$$

where $\boldsymbol{z}_{y,i}$ is the measurement vector from the left-hand side values of (12)–(14), and $\boldsymbol{z}_{y,i} \in \mathbb{R}^{(2M_i+3)\times 1}$; $\boldsymbol{h}(\boldsymbol{y}_i)$ as the measurement functions are calculated from the right-hand side of these formulas; $\boldsymbol{e}_{y,i}$ is the error vector in the coordinated phase, considering the measurement noises and estimation errors in local phases; $\boldsymbol{H}(\boldsymbol{y}_i)$ denotes the Jacobian matrix of $\boldsymbol{y}_i$, expressed in the block form as

$$\boldsymbol{H}(\boldsymbol{y}_i) = [\boldsymbol{H}^M, \boldsymbol{H}^S, \boldsymbol{H}^B] \quad (19)$$

where $\boldsymbol{H}^M$ and $\boldsymbol{H}^S$ denote the Jacobian matrices of the auxiliary measurement functions in (12) and (13), and $\boldsymbol{H}^M \in \mathbb{R}^{2M_i \times 2M_i}$, $\boldsymbol{H}^S \in \mathbb{R}^{1 \times 2M_i}$; $\boldsymbol{H}^B$ is the Jacobian matrix of (15) and (16). Table I lists the elements of these Jacobian matrices.

The measurement weight matrix $\boldsymbol{W}_{y,i}$ in (21) is expressed as

$$\boldsymbol{W}_{y,i} = diag(\boldsymbol{W}^M, \boldsymbol{W}^S, \boldsymbol{W}^B) \quad (20)$$

where $\boldsymbol{W}^M$ and $\boldsymbol{W}^S$ denote the measurement weights related to the boundary node, and their elements come from the corresponding diagonal elements of the covariance matrices in (5); $\boldsymbol{W}^B$ denotes the weight matrix on this boundary.

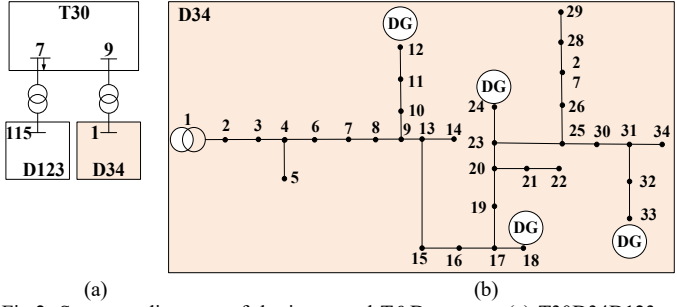

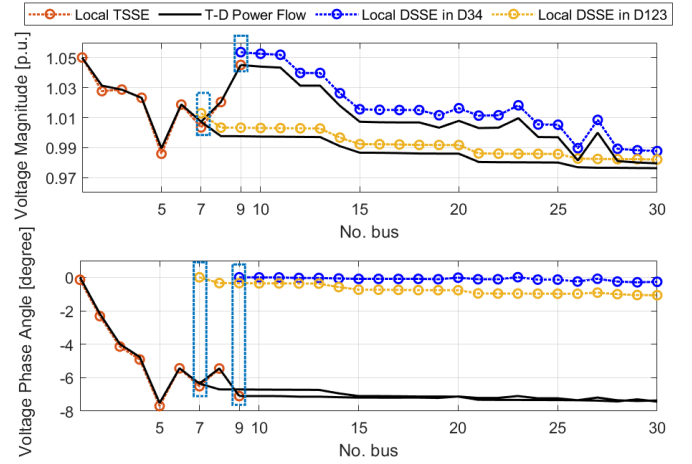

Fig.2. Structure diagrams of the integrated T&D systems (a) T30D34D123. (b) diagram of the IEEE 34-bus distribution system (D34 in T30D34D123)

Fig.3. Estimation results of parts of voltages by the local DSSE method

TABLE II
MEASUREMENT LOCATIONS IN DISTRIBUTION SYSTEMS

| Measurement Types | | Placement Locations | |
|---|---|---|---|
| | | 34-bus System | 123-bus System |
| SCADA | $|V|$ | 1, 11, 20, 27 | 11, 26, 99, 115 |
| | $P, Q$ | 1-2, 4-6, 16-17, 20-21 | 115-1, 1-7, 9-14, 15-16, 13-52, 18-35, 44-45, 57-60, 76-77, 86-87, 110-112 |
| Pseudo Meas. | | all load nodes and DG nodes | |

The estimation results for all the boundary buses are updated by (17)–(20) in the coordination phase. In summary, the proposed C-TDSE procedure is decomposed into three phases:

1) *Local Estimation Phase:* Obtain the local estimation results by solving (7) and (12). Then, store the state estimates at the nodes in $\mathcal{N}(i)$ and the corresponding elements of the gain matrices $\boldsymbol{G}_t$ and $\boldsymbol{G}_{d,i}$ for coordination.

2) *Coordination Phase:* (17) and (18) operate in each boundary system, and the initial values of states are chosen as the estimates in the local TSSE, *i.e.*, a hot start. The iterative tolerance at the coordination stage is set as $\varepsilon_c = 10^{-8}$.

The coordination terminates when $\Delta \boldsymbol{y}_i^{(\tau)}$ at all boundary nodes at iteration $\tau$ meets

$$\left\|\Delta \boldsymbol{y}_i^{(\tau)}\right\|_\infty = \left\|\begin{pmatrix}\Delta \boldsymbol{v}_n^{(\tau)} \\ \Delta \boldsymbol{\alpha}_n^{(\tau)}\end{pmatrix}\right\|_\infty < \varepsilon_c \quad i \in \{1,2,...,f\} \quad (21)$$

3) *Update Phase:* Feedback the updated boundary states to the local DSSE and TSSE algorithms, and the state variables are eventually refined based on the hot starts to achieve a more accurate estimate for the T&D system.

The convergence of the proposed method based on the heterogeneous decomposition of integrated T&D systems is guaranteed, and this proof can be found in [3].

## IV. NUMERICAL RESULTS

We test the proposed C-TDSE algorithm on an integrated T&D system. Illustrated as Fig. 2, the IEEE 30-bus transmission system connects with two distribution systems at buses 9 and 7, respectively. Also, these boundary nodes are connected to the substations of distribution networks. PMUs provide 15 voltage and 27 current measurements to achieve a high observability for this system. Gaussian noises with zero mean and maximum errors 1% of true values and 0.01 radians for magnitudes and phase angles are added to all the phasor measurements.

The buses located at substations of the IEEE 34- and 123-bus distribution feeders are numbered as 1 and 115 shown in Fig. 3, and see more details of these two systems in [9]. The installation details of these DGs can be referred to as [16]. All DGs are modeled as PQ buses with a constant power factor of 0.9, and the DG capacity is 600 kW. The measurement placement schemes are shown in Table III. The following conditions are applied to the maximum errors of measurements: SCADA systems measure voltage magnitudes and powers with the error margin up to 2%; pseudo-measurements provide the power measurements at loads and DGs, and the maximum errors are considered as 30% of the corresponding true values [16]. The master-slave-splitting power flow program of integrated T&D systems in [3] run for the accuracy check of state estimation. All test cases are performed for 200 times of Monte Carlo trials.

### A. Mismatch Reduction in Boundary Systems

The estimation results of the local TSSE and DSSE algorithms are compared with the true values from the coordinated power flow calculation of T&D systems. Limited to space, Fig. 3 shows the estimated voltages at the boundary nodes and parts of master and slave nodes in the test system. There are considerable mismatches of voltages at the boundary nodes. Due to lack of a global reference from PMUs, the local DSSE methods cannot correctly reflect the absolute phase angles of the nodal voltages in the slave systems.

Fig. 4(a) and 4(b) depict the convergence trend of the mismatches of powers at boundary nodes in the iterative process and the measurement residual in the coordination phase. Before the coordination, the mismatches of powers at the boundary nodes are shown at the leftmost ends of Fig. 4(a). These mismatches, which occur due to estimation errors on both levels of local estimation methods in this T&D system, are calculated by (14). It is observed that these residuals and mismatches rapidly decrease by the proposed algorithm, which is beneficial for TSOs to monitor the power flow directions and magnitudes at the boundaries systems.

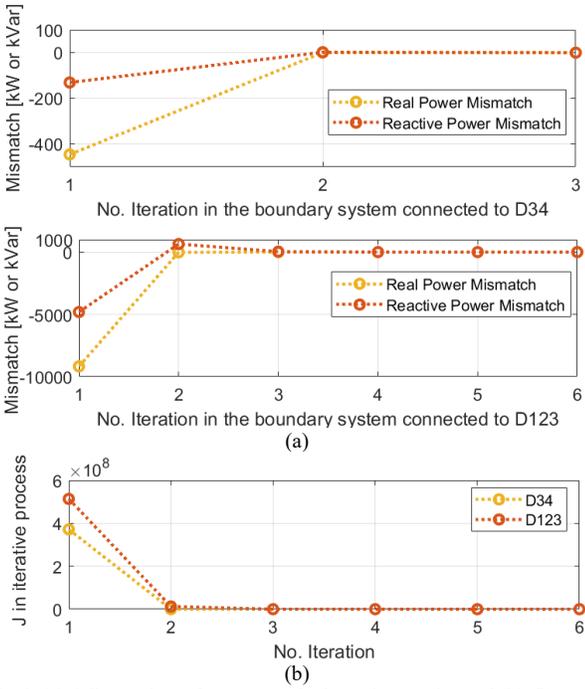

Fig.4. (a) Mismatches of powers at the boundary nodes and (b) $J(\boldsymbol{y}_i)$ of two boundary systems

TABLE III
RMSEs BEFORE AND AFTER COORDINATION

| Test Systems | | RMSEs at boundary nodes | Local TSSE | Local DSSE | C-TDSE |
|---|---|---|---|---|---|
| T30 | D34 | Mag. [%] | 0.1171 | 0.3380 | 0.1120 |
| | | Phase [degree] | 0.0641 | - | 0.0636 |
| | D123 | Mag. [%] | 0.2641 | 0.3425 | 0.1077 |
| | | Phase [degree] | 0.1565 | - | 0.0683 |

TABLE IV
COMPUTATION TIME AND NUMBER OF ITERATIONS

| Test System T30D34D123 | | Average Iter. | Time[ms] |
|---|---|---|---|
| TSSE | | - | 1.92 |
| Local DSSE | D34 | 3 | 136.7 |
| | D123 | 3 | 1638 |
| Coordinated Phase | D34 | 3 | 4.71 |
| | D123 | 5.78 | 4.28 |
| Update Phase | D34 | 2.14 | 109.7 |
| | D123 | 1.99 | 1269 |

*B. Estimation Accuracy and Computational Performance*

We discuss the estimation accuracy of the proposed method in integrated T&D systems. Root mean square errors (RMSEs) are used to evaluate the estimation performance [1] in all Monte Carlo simulations. The comparison of RMSEs of voltages at the boundary nodes is shown in Table III, and it should be noted that the RMSEs of phase angles are absolute errors. The proposed method improves the estimation accuracy at all the slave nodes. For instance, the RMSE of the voltage magnitudes at the boundary node in the 34-bus feeder decreases from 0.3380% to 0.1120%. It illustrates that the proposed method improves the overall estimation accuracy in the integrated T&D systems and accurately estimates the voltage phase angles at slave nodes with a global reference.

The computational performance of the proposed method, in terms of average iterations and CPU time, is investigated in comparison to the local TSSE and DSSE methods. Table IV lists the CPU time of these algorithms, where the tolerances are set as $\varepsilon_d = 10^{-4}$ and $\varepsilon_c = 10^{-8}$. The CPU time that the additional coordination and update phases take is 1.269 seconds. It is concluded that the proposed C-TDSE method has a fast converge speed and high computational efficiency.

## V. CONCLUSION

This paper proposes a decentralized and coordinated state estimation method in integrated T&D systems. The proposed algorithm enables more accurate monitoring of the integrated systems and reduces the mismatches of state variables at the boundaries of these systems, compared with individual TSSE and DSSE algorithms. In the case of lack of PMUs at the distribution level, this method accurately estimates the phase angles of the slave distribution systems in UTC.